\documentclass[12pt,fleqn]{article}
\RequirePackage{hyperref}
\usepackage{setspace}
\usepackage{amsmath,amsfonts,bm,amssymb}
\usepackage{amsthm}
\usepackage{ifpdf}
\usepackage{pdfsync}
\usepackage{eucal}
\usepackage{caption}
\usepackage[margin=2cm,a4paper]{geometry}
\usepackage[authoryear]{natbib}
\usepackage{url}\bibpunct{(}{)}{,}{a}{,}{,}
\usepackage{multirow}
\usepackage{rotating}
\usepackage{graphicx}
\usepackage{grffile}
\usepackage{color}
\usepackage[affil-it]{authblk}

\usepackage[authoryear]{natbib}
\usepackage{url}
\bibpunct{(}{)}{,}{a}{,}{,}

\renewcommand{\leq}{\leqslant}

\renewcommand{\geq}{\geqslant}
\newcommand{\hrc}{\hat\rho^c}
\newcommand{\htc}{\hat\theta^c}
\newcommand{\boot}{\mathrm{b}}

\newcommand{\remark}[1]{\marginpar{\scriptsize#1}}
\renewcommand{\remark}[1]{\marginpar{}}  

\long\def\symbolfootnote[#1]#2{\begingroup\def\thefootnote{\fnsymbol{footnote}}\footnote[#1]{#2}\endgroup}
\newcommand{\note}[1]{\footnote{#1}}

\newtheorem{theorem}{Theorem}
\newtheorem{lemma}[theorem]{Lemma}
\newtheorem{corollary}[theorem]{Corollary}
\newtheorem{preposition}[theorem]{Preposition}

\makeatletter
\def\@maketitle{%
  \newpage
  \null
  \vskip 2em%
  \begin{center}%
  \let \footnote \thanks
    {\Large\bfseries \@title \par}%
    \vskip 1.5em%
    {\normalsize
      \lineskip .5em%
      \begin{tabular}[t]{c}%
        \@author
      \end{tabular}\par}%
    \vskip 1em%
    {\normalsize \@date}%
  \end{center}%
  \par
  \vskip 1.5em}
\makeatother

\begin{document}

\title{Value Matters:\\
Predictability of Stock Index Returns}

\author{Natascia Angelini%
  \thanks{Electronic address: \texttt{natascia.angelini@unibo.it}}}
  \affil{MEDAlics, Dante Alighieri University 
    
    Via del Torrione 95, 89125 Reggio Calabria, Italy 

    School of Economics, Management and Statistics, University of Bologna

    Via Angher\`a 22, 40127 Rimini, Italy
  }

\author{Giacomo Bormetti%
  \thanks{Electronic address: \texttt{giacomo.bormetti@sns.it}}}
  \affil{Scuola Normale Superiore

  Piazza dei Cavalieri 7, 56126 Pisa, Italy
  }

\author{Stefano Marmi%
  \thanks{Electronic address: \texttt{stefano.marmi@sns.it}}}
  \affil{Scuola Normale Superiore

  Piazza dei Cavalieri 7, 56126 Pisa, Italy}

\author{Franco Nardini%
  \thanks{Electronic address: \texttt{franco.nardini@unibo.it}; Corresponding author, mobile phone: +39 335 8024631}}
  \affil{Department of Mathematics, University of Bologna,

  Viale Filopanti 5, 40126 Bologna, Italy}

\date{}

\maketitle

\vspace{0.3cm}

\begin{abstract}
  We present a simple dynamical model of stock index returns which is grounded on the ability of the Cyclically Adjusted Price Earning (CAPE) valuation ratio devised by Robert Shiller to predict long-horizon performances of the market. More precisely, we discuss a discrete time dynamics in which the return growth depends on three components: i) a momentum component, naturally justified in terms of agents' belief that expected returns are higher in bullish markets than in bearish ones, ii) a fundamental component proportional to the logarithmic CAPE at time zero. The initial value of the ratio determines the reference growth level, from which the actual stock price may deviate as an effect of random external disturbances, and iii) a driving component which ensures the diffusive behaviour of stock prices. Under these assumptions, we prove that for a sufficiently large horizon the expected rate of return and the expected gross return are linear in the initial logarithmic CAPE, and their variance goes to zero with a rate of convergence consistent with the diffusive behaviour. Eventually this means that the momentum component may generate bubbles and crashes in the short and medium run, nevertheless the valuation ratio remains a good reference point of future long-run returns.
\end{abstract}


\section{Introduction}

A question that has been asked with a certain regularity since the establishment  of the first stock index is the following, see~\citet{Myers_Swaminathan:1999}: is it possible to calculate ``the Intrinsic Value of the Dow''?  Or, more generally: what is the intrinsic value of Wall Street? A first commonsense answer is that this value may be calculated by summing ``the intrinsic value'' of each quoted company. At the beginning of the past century, see~\citet{Williams:1938, Graham_Spencer:1998}, it was well known that the latter depends on the expected future earnings and on the balance sheet, but unfortunately there was no easy formula, which yielded even a rough approximation. Therefore 
aggregating all expectations for all companies quoted in the market soon appeared to be a challenging task.

Between the Fifties and the Sixties the researchers' attitude towards the problem changed radically. Modern portfolio theory and the statistical evidence that stock prices follow random walks, see~\citet{Mandelbrot:1963, Fama:1965}, has led to the Efficient Market Hypothesis, see~\citet{Samuelson:1965, Fama:1970}. According to this theory, financial markets are ``informationally efficient'', that is, one can not achieve returns in excess of average market returns on a risk-adjusted basis, given the information available at the time the investment is made. If this were the case ``the intrinsic value'' of a stock would be simply its market price and the problem would ultimately be non existent. The bubble burst in 1987 and the exceptional price boom in the late '90 lead more and more researcher to cast doubts on the hypothesis' truth. At the very beginning of 2000 Robert Shiller wrote: ``we do not know whether the market level makes any sense, or whether they are indeed the result of some human tendency that might be called irrational exuberance'',~\citet{Shiller:2000}, which is apparently at odds with the assumption that market prices are always fair prices. Shiller's skeptical attitude was anticipated in one of the most famous books ever written on the stock market: in 1934, Graham and Dodd strongly advocated the fundamental approach to investment valuation and recommended to ``shift(s) the original point of departure, or basis of computation, from the current earnings to the average earnings, which should cover a period of not less than five years, and preferably seven to ten years''~\citet{Graham_Dodd_Cottle:1962}. Since 1988 Campbell and Shiller have been  focusing their attention on average earnings as predictor of future dividends~\citet{Campbell_Shiller:1988a}~\note{``Our results indicate that a long moving average of real earnings helps to forecast future real dividends. The ratio of this earnings variable to the current stock price is a powerful predictor of the return on stock, particularly when the return is measured over several years''.} and future stock prices~\citet{Campbell_Shiller:1998}~\note{The price-smoothed earnings ratio has little ability to predict future growth in smoothed earnings; the R squared statistics are less than 4\% over one year and over ten years. The ratio is a good forecaster of ten-year growth in stock prices, with an R squared statistic of 37\%.};~\citet{Shiller:2000} proposed an innovative test of the appropriateness of prices in the stock market: the Cyclically Adjusted Price Earning ratio, and showed that it is a powerful predictor of future long run performances of the market; the performance of the test is quite satisfactory in the case of the US market from the end of 19$^\mathrm{th}$ century up to today. In the same years, again elaborating on~\citet{Graham_Dodd_Cottle:1962},~\citet{Lander_etal:1997} suggested an alternative but similar criterion based on the assumption ``that many investors are constantly making a choice between stock and bond purchases; as the yield on bonds advances, they would be expected to demand a correspondingly higher return on stocks, and conversely as bond yields decline~\citet[pag. 510]{Graham_Dodd_Cottle:1962}''; their ``approach considers whether stocks are appropriately valued relative to analysts' perceptions of future earnings and yields on alternative investments'' such as high grade corporate bonds (for a comparison between the two approaches see~\citet{Weigand_Irons:2006}).

The aim of this paper is to provide a theoretical framework which explains how future stock index returns depend on the CAPE ratio. This is achieved by constructing a model.
What we require of our model is similar to what Chiarella required in~\citet[see pag.~102]{Chiarella:1992}

``Firstly we require that the model generate a significant transitory component around the equilibrium which reflects the rationally expected
value of the asset. Secondly the model must allow for the incorporation of chartists, a group which bases its market actions on an analysis of past
trends. Since this group seems to be an important part of real markets it is important to determine what effect its activity has on the behaviour of asset
prices and whether the behaviour of a model incorporating chartists comes closer to explaining some of the empirical results cited earlier''. 

We choose, however, a different approach and we work at an aggregate level  instead of explicitly modelling two different types of agents.  
We introduce a moment component in a ``simple error-correction model that predicts the return of the S\&P based on the deviations 
from a presumed equilibrium between \ldots earnings yields''~\citet[see pag.~3]{Lander_etal:1997}
and a long run target yield rate. The return growth depends on three components
\begin{itemize}
  \item[a)] a momentum component, naturally justified in terms of agents' expectation that expected returns are higher in bullish markets than in bearish ones;
  \item[b)] a fundamental component, proportional to a function of the level of the logarithmic averaged earnings-to-price ratio (for brevity log EP ratio) 
    at time zero; the initial time value of log EP ratio determines the reference growth level, from which the actual stock price may deviate as an effect of random external disturbances;
  \item[c)] a driving component ensuring the diffusive behaviour of stock prices.
\end{itemize}
Under these assumptions, we are able to prove that, if we consider a sufficiently large number of periods, the expected rate of return and the expected gross return are linear in the initial time value of log EP, and their variance converges to zero with rate of convergence equal to minus one. This means that, in our model, the stock prices dynamics may generate bubbles and crashes in the short and medium run, nevertheless the averaged earnings-price ratio is a good predictor of future long-run returns, as claimed by~\citet{Campbell_Shiller:1988a, Shiller:2000, Lander_etal:1997}. The result holds for both returns and gross returns; in the latter case we assume that the log dividend-to-price ratio follows a stationary stochastic process as in~\citet{Campbell_Shiller:1988a,Campbell_Shiller:1988b}. Relaxing this assumption~\citet{Engsted_Pedersen_Tanggaard:2012} 
have recently investigated a model where explosive bubbles may arise, while~\citet{Campbell:2008} has set forth an innovative approach to estimate the equity risk premium.

In section~\ref{sec:econframe} we review the econometric issues arising in when a rate of return is regressed on a lagged stochastic regressor where the regression disturbance is correlated with the regressor's innovation. In section~\ref{sec:data} we describe the data set we are going to use in the estimation of the model. Section~\ref{sec:dynsystem} sets up the model and derives the theoretical results. In section~\ref{sec:estimation} we estimate the parameters of our model for both the Standard and Poor Composite Stock Price Index and the NYSE/AMEX value weighted index, and we present the results of a Monte Carlo simulation, which confirms the ability of the proposed model to capture the relevant features of the historical time series. Section~\ref{sec:conclusions} concludes. Technical proofs are gathered in the appendix.

\section{Econometric framework}\label{sec:econframe}

The approach that we discuss in this paper is based on the advances achieved in recent years by different authors to deal with persistent regressions. Although the methodology that we employ is of general applicability, we detail it with a specific focus on the topic of index returns predictability with respect to valuation ratios.\\

\noindent We refer to the real (inflation adjusted) price of the stock index, measured at the beginning of time period $t$, with $P_t$, while the notation $D_t$ corresponds to the real dividend paid on the portfolio during the period between $t$ and $t+1$. Accordingly, we write the real log gross return on the index held from the beginning of time $t$ and the beginning of time $t+1$ as
\begin{equation*}
  H_t = \log{\left(P_{t+1}+D_t\right)}-\log{P_t}\, .
\end{equation*}
We provide a description of the return dynamics on a monthly basis. Therefore, the notation $t+1$ refers to the time instant $t$ increased by one month and the real gross yield over a period of length $h$ months corresponds to
\begin{equation}
  y_{t,h} = \frac{1}{h}\sum_{i=0}^{h-1} H_{t+i}\, .
  \label{eq:yield}
\end{equation}
We also introduce the index log price $p_t=\log{P_{t}}$, in terms of which the gross yield can be rewritten as 
\begin{equation}
  y_{t,h} = \frac{1}{h}\sum_{i=0}^{h-1} \left (p_{t+i+1} - p_{t+i}\right)+\frac{1}{h}\sum_{i=0}^{h-1} \log{\left(1+\frac{D_{t+i}}{P_{t+1+i}}\right)}\, ,
\end{equation}
where the first term on the right hand side is equivalent to $(p_{t+h} - p_t)/h$. The dependent variable that we study throughout the paper is the gross return of the stock index, while the regressor that we consider is the log Earnings-to-Price ratio (log EP) $x_t \doteq \log{\langle e \rangle_t^{10}} - p_t$. The quantity $\langle e \rangle_t^{10}$ refers to a moving average of real earnings over a time window of ten years. At variance with~\cite{Campbell_Shiller:1988b}, where the analysis is based on geometrical averages, we employ arithmetical averages. The very reason is that in this way our study can be extended to the case of negative earnings. Indeed, while it is quite unrealistic that an average over ten years results in a negative value, however it is not so uncommon to deal with spot negative earnings. In this latter case the definition of $\log{e_t}$ would be problematic. Moreover, we expect that switching from one average to the other does not significantly change the overall picture. The use of an average of earnings in computing the ratios has been strongly pushed by the literature in recognition of the cyclical variability of earnings. In~\citet{Graham_Dodd_Cottle:1962} the authors recommend an approach that ``Shifts the original point of departure, or basis of computation, from the current earnings to the average earnings, which should cover a period of not less than five years, and preferably seven to ten years.''\\

\noindent We assume that the return $y_t$\footnote{To ease the notation from the remaining of this section we express the log return $y_{t-1,1}$ as $y_t$.} and the predictor satisfy the relations which follow
\begin{eqnarray}
  y_t  &=& \alpha + \beta ~x_{t-1} + u_t\,,\label{eq:regression}\\
  x_t &=& \theta + \rho  ~x_{t-1} + v_t\,,\label{eq:AR1}
\end{eqnarray}
where the errors $\left(u_t, v_t\right)$ are serially independent and identically distributed as bivariate normal, with zero mean and covariance matrix $\Sigma$. We follow the same formal model of~\cite{Stambaugh:1999} where the time series of returns $\{y_t\}$ for $t=1,\ldots,n$ is to be predicted by a scalar first-order autoregressive time series $\{x_t\}$ for $t=0,\ldots,n-1$. When the regressions of the above type involve financial ratios the Ordinary Least-Squares (OLS) estimator for $\alpha$ and $\beta$ is biased in finite samples. This problem is pointed out by many authors, see for instance~\cite{Mankiw_Shapiro:1986,Stambaugh:1986}, and is analyzed by~\cite{Stambaugh:1999} who develops an expression for the bias of the estimated prediction coefficient. In Stambaugh's words ``An assumption that typically fails to hold (\dots) is that $u_t$ has zero expectation conditional on $\{\ldots,x_{t-1}, x_t, x_{t+1}, \ldots\}$, and this is the assumption used to obtain finite-sample results in the standard setting.'' Stated in different words, the behaviour of $u_t$ is mainly determined by the values $p_{t-1}$ and $p_t$ so $\mathbb{E}(u_t | x_t, x_{t-1})\neq 0$. Moreover, the dynamics of $x_t$ is strongly persistent and largely influenced by the behaviour of $p_t$. When the log price $p_{t-1}$ increases, the idiosyncratic term $u_t$ increases too, while $v_t$ decreases; the converse is true when the price decreases. Empirically the value of $\rho$ is relatively close to one and as an overall effect this induces a strong anti-correlation between $u_t$ and $v_t$. When both the endogeneity of the regression and the persistence of $x_t$ are taken into account the high statistical significance of the standard OLS estimator, $\hat\beta^{\text{OLS}}$, found in~\cite{Campbell_Shiller:1987,Campbell_Shiller:1988a,Campbell_Shiller:1988b} is to be questioned. However, starting from~\cite{Stambaugh:1999}, \cite{Lewellen:2004} finds some evidence for predictability with valuation ratios. This result relies on the upper bound for the bias in $\hat\beta$ which derives from the conservative assumption that the autoregressive coefficient is very close to one. We follow the alternative approach discussed in~\cite{Amihud_Hurvich:2004}. Replacing the predictive regression~(\ref{eq:regression}) with the augmented regression 
\begin{equation}
  y_t = \alpha^c + \beta^c x_{t-1}+ \phi^c v_t^{c} + e_t\,,
  \label{eq:augmented}
\end{equation}
where  
\begin{equation}
  v_t^c = x_t - \htc - \hrc x_{t-1}\,,
  \label{eq:proxy}
\end{equation}
it is possible to show that the OLS estimator of $\beta^c$ is a low-biased estimator of the true parameter $\beta$ appearing in~(\ref{eq:regression}). The quantities $\htc$ and $\hrc$ needed to define the regressor $v_t^c$ correspond to reduced-bias estimators of $\theta$ and $\rho$, respectively. A possible choice for $\hrc$ corresponds to the Kendall's low-bias estimator of the AR(1) coefficient discussed in~\cite{Kendall:1954}. At variance we employ the second-order bias corrected estimator proposed in~\cite{Amihud_Hurvich:2004} which reads
\begin{equation}
  \hrc = \hat\rho^{\text{OLS}} + \frac{1+3\hat\rho^{\text{OLS}}}{n} + 3 \frac{1+3\hat\rho^{\text{OLS}}}{n^2}\,,
  \label{eq:hrc}
\end{equation}
where $\hat\rho^{\text{OLS}}$ is the standard OLS estimator. Coherently, the following relation holds
\begin{equation}
  \htc = (1 - \hrc) \frac{\sum_{t=0}^{n-1} x_t}{n}\,,
  \label{eq:htc}
\end{equation}
with $n$ equal to the number of available observations. The statistical significance of $\beta$ can be tested by means of a low-bias finite sample approximation of the standard error of $\hat\beta^c$ which deviates from the error provided by the OLS regression. The reason for this deviation is that the OLS estimate fails to take into account the additional variability of $\hat\beta^c$ due to the variability of $\hrc$. Unfortunately, within the same framework it seems not possible to derive an expression for the error associated with $\alpha$. Testing their approach on a data sample of monthly records from the NYSE inflation-adjusted value-weighted index ranging from May 1963 to December 1994, Amihud and Hurvich find a highly statistically significant predictability for dividend yields. In the present paper we perform the estimation of the $\alpha$ and $\beta$ coefficients exploiting the same methodology. However, as far as testing is concerned, we do not exclusively rely on the $t$-statistic but we resort to a bootstrap approach which does not rely on any distributional assumption for the residual components $u_t$ and $v_t$. As we will show in section~\ref{sec:estimation} we find coherent results which support the forecasting ability of the logarithmic CAPE for both the Standard and Poor and NYSE/AMEX market returns.\\

\noindent Before we move to the description of the bootstrap algorithm, it is worth to mention some interesting alternative results dealing with the subtle issue of predictability with valuation ratios. To the best of our knowledge, the most recent advance corresponds to~\cite{Hjalmarsson:2011} which is inspired by the testing framework discussed in~\cite{Campbell_Yogo:2005,Campbell_Yogo:2006}. Both approaches are grounded on the quasi-unit root approximation for the AR(1) coefficient of equation~\ref{eq:AR1}, but while the latter focuses on short-run horizons the former extends the results to arbitrary horizons. The extension to the long-run predictability involves a new element that contributes to increase the complexity of the econometric issue. Indeed, long-run performances can be evaluated only aggregating returns over smaller time scales. Given the intrinsic scarcity of data points, the sample can be enlarged only resorting to overlapping realizations and inducing sizable serial correlation among the residual terms. This side-effect can be managed by means of auto-correlation robust estimators, for instance those discussed in~\cite{Hansen_Hodrick:1980,Newey_West:1987}, however, when the correlation is particularly strong, these methods show poor small sample properties as discussed in~\cite{Hodrick:1992,Nelson_Kim:1993}. On the other hand the new asymptotic results developed in~\cite{Hjalmarsson:2011} are  based on the scaling properties of the $t$-statistics and are tested on a sub sample of the data used in~\cite{Campbell_Yogo:2006} providing consistent results and evidence of the predictive power of the logarithmic CAPE.\\  

\noindent The bootstrap in the context of returns predictability is discussed in various papers,~\cite{Kilian:1999,Goyal_Santa-Clara:2003,Goyal_Welch:2008}, and more specifically for regressions based on valuation ratios in~\cite{Maio:2012}. The testing environment that we design goes through the following steps:
\begin{enumerate}
  \item[i)] We estimate the bias-corrected $\hrc$ and $\htc$ by means of~(\ref{eq:hrc}) and~(\ref{eq:htc}). Using both values we compute~(\ref{eq:proxy}) and perform the augmented regression~(\ref{eq:augmented}). We store the values $\hat\alpha^c$ and $\hat\beta^c$ that we find via OLS. 
  \item[ii)] We consider the regression~(\ref{eq:regression}) and~(\ref{eq:AR1}) under the null hypothesis $\beta=0$ and we estimate the coefficients via OLS. We save the values $\hat\alpha$, $\hat\theta$, and $\hat\rho$ together with the sequence of bivariate residuals $(u_t, v_t)$ for $t=1,\ldots,n$. 
  \item[iii)] We sample $\boot=1,\ldots,10^4$ bivariate sequences $(u_t^\boot, v_t^{\boot})$ with $t=s_1^\boot,\ldots,s_n^\boot$. The time indices $s_1^\boot,\ldots,s_n^\boot$ are sampled with replacement from the original set of indices $t=1,\ldots,n$. 
  \item[iv)] For each replication $\boot=1,\ldots,10^4$, we compute a bootstrap sample by imposing the null hypothesis 
    \[
	y_t^\boot = \hat\alpha + u_t^{\boot}\,,
    \]
    \[
      x_t^\boot = \hat\theta + \hat\rho x_{t-1}^\boot + v_t^{\boot}\,.
    \]
    The initial point $x_0^\boot$ is drawn uniformly from the set $\{x_t\}_{t=0,\ldots,n-1}$.
  \item[v)] For each bootstrap sample we estimate the bias-corrected $\hat\theta^{c,\boot}$ and $\hat\rho^{c,\boot}$, we compute the proxy for the error terms $v_t^{c,\boot}$ as
    \[
      v_t^{c,\boot} =  x_{t}^\boot - \hat\theta^{c,\boot} - \hat\rho^{c,\boot} x_{t-1}^\boot\,,
    \]
    and we estimate the following augmented regression via OLS
    \[
      y_t^\boot = \alpha^{c,\boot} + \beta^{c,\boot} x_{t-1}^\boot+ \phi^{c,\boot} v_t^{c,\boot} + e_t^\boot\,.
    \]
    Finally we obtain a sample $\left\{\hat\beta^{c,\boot}\right\}$ for $\boot=1,\ldots,10^4$.
  \item[vi)] The one sided $p$-value associated with the prediction coefficient is calculated as 
    \[
      p-\text{value}=\frac{\#\{\hat\beta^{c,\boot}\geq\hat\beta^c\}}{10^4}\,,
    \]
    where the numerator on the right hand side denotes the number of bootstrap estimates higher than the original value, and we implicitly assume that $\hat\beta^c > 0$.
\end{enumerate}

\section{Data set}\label{sec:data}
The data set analyzed in this paper consists of records on a monthly basis of prices, earnings, and dividends for two stock price indexes for the US market. The first is the Standard and Poor Composite Stock Price Index (S\&P) data discussed in~\citet{Campbell_Shiller:1987,Campbell_Shiller:1988a,Campbell_Shiller:1988b}, and freely available from Robert J. Shiller's webpage \emph{http://www.econ.yale.edu/}. These time series cover the entire period from January 1871 until December 2012. The second is NYSE/AMEX value weighted index data (1926-1994) from the Center for Research in Security Prices (CRSP) available from J. Campbell home page \emph{http://scholar.harvard.edu/campbell/data}. Since earnings data are not available for CRSP series we use the corresponding S\&P earnings-to-price ratios from Shiller. Figure 1 in~\citet{Campbell_Yogo:2006} compares the time series of the log dividend-to-price ratio for the NYSE/AMEX value-weighted index and the log smoothed earnings-price for the S\&P and provides support for our choice.

\section{The model}\label{sec:dynsystem}
In the current section we characterise a simple dynamical system driven both by a value investment and a momentum component. In the first instance we consider only the role played by stock prices, but later we take into account gross returns and the effect of dividends. Our model rests on the two assumptions which follow.\\

\noindent\textbf{Assumption 1}
\textit{
  Financial returns are naturally driven by the growth rate $\mu_t$, and by an ancillary process $\xi_t$ satisfying 
  \begin{equation}
    \xi_{t+1} = \xi_{t}+\frac{\kappa}{1-\gamma}\sigma_p W_{t}^p\, ,
  \end{equation}
  with initial time condition $\xi_0=0$. The $\xi_t$ process ensures the diffusive behaviour of stock prices. 
  As will be clear later, the positive parameter $\sigma_p$ corresponds to the long run returns volatility, 
  while the prefactor depending on $\kappa>0$ and $0<\gamma<1$ has been added for future convenience.
  The $W_t^p$'s for $t = 0,\ldots,h$ are independent identically distributed (i.i.d.) standard Gaussian increments. 
}\\

\noindent\textbf{Assumption 2} 
\textit{The dynamics of log returns' growth rate is given in terms of the difference equation
  \begin{equation}
    \mu_{t+1}  = \gamma \mu _{t} + \kappa \left[\log \langle e\rangle_{t}^{10} - p_t + \mathcal{H} + g\mathcal{F}\left(\log \langle e\rangle_{0}^{10}-p_0\right)t 
      \right] + \sigma_\mu W_t^\mu
    \end{equation}
    where the parameter $\sigma_\mu$ is a positive volatility constant, and the $W_t^\mu$'s for $t = 0,\ldots,h$ are {i.i.d.} standard Gaussian increments,
not dependent on the $W_t^p$'s. 
The return growth $\mu_t$ depends on its own value in the previous period, with $\gamma$ agents' sensitivity to market trend.
This momentum effect can be naturally justified in terms of agents' expectation that returns are higher in bullish markets than in bearish ones.
The ``fundamental'' component is proportional to a function $\mathcal{F}$ of the 
level of the logarithmic averaged earnings-to-price ratio at time zero, 
where the proportionality constant is given by the rate of earnings growth $g$, implicitly defined by $\langle e \rangle_t^{10}=\langle e \rangle_0^{10} \exp{(g t)}$.
This component grows linearly with time, but we also allow for a correction equal to $\mathcal{H}$, possibly function of the initial time log EP ratio.
The initial time value of log EP determines the reference growth level, from which the actual stock price may deviate as an effect of random external disturbances.
Investors reallocate assets in response to this disequilibrium causing stock prices to move in the direction that reduces the deviation.
The full adjustment is not immediate but it takes a typical time $\kappa^{-1}$ the price to mean revert to the fundamental level.
}\\  
In conclusion, our model states the dynamics of log prices can be fully specified by the linear stochastic difference system 
\begin{equation}
	\left\{ 
	\begin{array}{rl}
		p_{t+1}   & = p_{t} + \mu _{t} + \xi_{t} \\ 
		\mu_{t+1} & = \gamma \mu _{t} + \kappa \left(\log \langle e\rangle_{t}^{10} - p_t + \mathcal{H} + g\mathcal{F}t\right) + \sigma_\mu W_t^\mu \\ 
		\xi_{t+1} & = \xi_{t} + \frac{\kappa}{1-\gamma}\sigma_p W_{t}^p
	\end{array}
	\right.\, ,
	\label{eq:linear_SDS}
\end{equation}
with $p_0=\log P_0$ and $\mu_0$ initial time conditions. By a further change of variable
$Y_t=p_t-\log \langle e\rangle_{0}^{10}$, the expected values $\mathbb{E}_0[Y_t]$,  $\mathbb{E}_0[\mu_t]$, $\mathbb{E}_0[\xi_t]$
conditional at the information available at time $0$ evolve according to the linear first order difference system
\begin{equation*}
	\left\{ 
	\begin{array}{ll}
		\mathbb{E}_0[Y_{t+1}]   & =        \mathbb{E}_0[Y_{t}] +        \mathbb{E}_0[\mu_{t}]   + \mathbb{E}_0[\xi_t]\\ 
		\mathbb{E}_0[\mu_{t+1}] & = -\kappa \mathbb{E}_0[Y_{t}] + \gamma \mathbb{E}_0[\mu _{t}] + \kappa\mathcal{H}  + \kappa g(1+\mathcal{F}) t\\ 
		\mathbb{E}_0[\xi_{t+1}] & =                                     \mathbb{E}_0[\xi_{t}]
	\end{array}
	\right.\, , 
\end{equation*}
with the initial conditions
\begin{equation*}
	\left\{ 
	\begin{array}{ll}
		\mathbb{E}_0[Y_{0}]    & =  p_0-\log \langle e\rangle_{0}^{10} \\
		\mathbb{E}_0[\mu_{0}]  & = \mu_0 \\ 
		\mathbb{E}_0[\xi_{0} ] & = 0 
	\end{array}
	\right.\, .
\end{equation*}
\begin{lemma}
	The expected rate of return is linear in $\mathcal{F}(\log \langle e\rangle_0^{10}-p_0)$, 
	provided we consider a sufficiently large number of periods $h$  
	\begin{equation}
		\mathbb{E}_{0}\left[ \frac{1}{h}\log \frac{P_{0+h}}{P_{0}}  \right] = g (1+\mathcal{F}) + O(\frac{1}{h}) \, . 
		\label{eq:th_mean}
	\end{equation}
	Moreover its variance converges to zero with rate of convergence equal to minus one
	\begin{equation}
		Var_{0}\left[ \frac{1}{h}\log \frac{P_{0+h}}{P_{0}} \right] = \frac{\sigma_p^2}{h} + o(\frac{1}{h})\, .
		\label{eq:th_var}
	\end{equation}
\end{lemma}
A posteriori it is clear why we fixed the proportionality constant of the fundamental component equal to $g$. Indeed, in case $g$ were zero not only averaged earnings but also stock returns would grow sub-exponentially. The variance of $Y_h$ increases linearly with $h$, the diffusive behaviour of stock returns that any realistic model should satisfy in a first instance. Finer effects, like possible return non zero autocorrelations induced by business cycles over long horizons could be included in our model modifying the  covariance structure of the $\{W_t^{p}\}$ noise. In this respect it is worth to remark that none of our results crucially rest on the Gaussian assumption for the noise increments.\\
From previous Lemma the result that we state below follows.
\begin{corollary} 
	If $\mathcal{F}$ is linear in $\log \langle e\rangle_0^{10}-p_0$, than the model~(\ref{eq:linear_SDS}) is able to reproduce the linear scaling of returns with the initial 
	log EP ratio on the long run.
        \label{cor:Flinear}
\end{corollary}
The next step forward is to add a dividend components to our stock returns model. In this respect we mainly follow the proposal discussed 
in~\cite{Campbell_Shiller:1988a,Campbell_Shiller:1988b}, where it is argued that the log dividend-to-price ratio follows a stationary stochastic process, and 
the fixed mean of $\log {D_t} - \log {P_t}$, $\log \mathcal{G}$, can be used as an expansion point
\begin{equation}
	\Delta(d_{t-1}-p_{t}) = -\theta (d_{t-1}-p_t-\log\mathcal{G}) + \sigma_\mathrm{d} W_t^{\mathrm{d}}\, , 
	\label{eq:logDP}
\end{equation}
with $d_t=\log D_t$, $\sigma_\mathrm{d}>0$, and $\{W_t^\mathrm{d}\}$ for $t=1,\ldots,h$ {i.i.d.} standard Gaussian variates. At variance with the proposal
of Campbell and Shiller, we also assume that $\mathcal{G}$ can possibly depend on the log EP at time zero, 
i.e. $\mathcal{G}=\mathcal{G}(\log \langle e\rangle_0^{10}-p_0)$. 
\begin{lemma}
	In the limit $h\gg 1$ the quantity $\frac{1}{h}\sum_{i=0}^{h-1} \log{\left(1+\frac{D_{i}}{P_{i+1}}\right)}$
	contributes to the rate of log gross returns proportionally to $\mathcal{G}$, and its variance converges to zero with rate minus one.
\end{lemma}
\begin{corollary} 
	If $\mathcal{G}$ is linear in $\log \langle e\rangle_0^{10}-p_0$, than equation~(\ref{eq:logDP}) is able to reproduce the linear scaling of dividends' contribution to the yield 
	with the initial log EP ratio on the long run.
        \label{cor:Glinear}
\end{corollary}
We are now ready to state the main theoretical result of this paper.
\begin{preposition}
	The expected gross yield~(\ref{eq:yield}) is linear in $\mathcal{F}$ and $\mathcal{G}$, provided we consider sufficiently long horizons
	\begin{equation}
		\mathbb{E}_0[y_h] \simeq g (1+\mathcal{F}) + \mathcal{G} + O\left(\frac{1}{h}\right)\, .
		\label{eq:mainprepo}
	\end{equation}
\end{preposition}
The proof is an immediate consequence of the Corollaries~\ref{cor:Flinear} and~\ref{cor:Glinear}. Interestingly from the proofs of previous Lemmas 
given in Appendix~(\ref{app:proof}) we can explicitely
identify the contribution to the long term yield whose scaling over time is more persistent, and which ultimately determines its convergence towards the 
limit  $g (1+\mathcal{F}) + \mathcal{G}$. The expression of the leading correction proportional to one over $h$ reads
\begin{equation}
	\mathcal{H} - g (1+\mathcal{F})\left[1+\kappa\frac{1-\lambda_-\lambda_+}{(1-\lambda_-)^2(1-\lambda_+)^2}\right]+\log EP_0
	+\mathcal{G}\left(1-\frac{1}{\theta}\right)(\log DP_0-\log\mathcal{G})\,.
	\label{eq:leading}
\end{equation}
From equation~(\ref{eq:expected_dX}) we know that $(1-\theta)^t$ plays the role of a damping function. Assuming $\theta\ll 1$, it reduces to $\exp(-t/\tau_\theta)$
where the quantity $\tau_\theta=1/\theta$ plays the role of the typical time scale of the mean-reverting process. We reasonably expect that the last term in the 
expression~(\ref{eq:leading}) does not contribute too much to the leading correction, since we neither expect an extremely large time scale for the process 
nor an extreme discrepancy between $\log DP_0$ and $\log\mathcal{G}$. Another interesting point to note is that the choice of $\mu_0$ plays a minor role 
in relation to the speed of convergence of the process to the long term expected value. 
Indeed, $\mu_0$ does not appear in expression~(\ref{eq:leading}), while from~(\ref{eq:expected_Y}) we know that its effect is exponentially damped by $\lambda_-^h$ and $\lambda_+^h$. 
Being $0<\lambda_-<\lambda_+<1$, the dominating contribution for $h\gg 1$ is given by $\lambda_+^h$, implying a typical damping scale equal to $-(\log \lambda_+)^{-1}$.

\section{Parameters estimation}\label{sec:estimation}
Corollaries~\ref{cor:Flinear} and~\ref{cor:Glinear} state that the linear scaling of long term yield with the level of log EP ratio can be reproduced by our dynamical model if we assume for the quantities $g(1+\mathcal{F})$ and $\mathcal{G}$ the form $g(1+\mathcal{F})=\alpha_\mathcal{F}+\beta_\mathcal{F}\log\mathrm{EP}_0$ and $\mathcal{G}=\alpha_\mathcal{G}+\beta_\mathcal{G}\log\mathrm{EP}_0$. Following the same approach of~\cite{Amihud_Hurvich:2004}, we estimate the values of $\alpha_{\mathcal{F}}$, $\beta_{\mathcal{F}}$ from the regression of log returns over one year horizon on the initial log EP. The quantities labelled with $_\mathcal{G}$ are obtained in the same way after regressing the dividends' contribution to the yield on the same regressors. The results that we find are reported in Table~\ref{tab:mean_alfabeta} for both the S\&P and NYSE/AMEX indices. We also report the prediction coefficients of the gross returns on the initial logarithmic CAPE, $\alpha$ and $\beta$. The statistical significance of the $\beta$ coefficients is assessed by means of the reduced-biased standard errors discussed in~\cite{Amihud_Hurvich:2004} and according to the bootstrap approach devised in section~\ref{sec:econframe}. If we compute the $t$-statistics for the $\beta$ parameters we find $t=2.29$ and $t=2.52$ for the S\&P and CRSP time series respectively, which imply a statistical significant evidence of predictability. The predictive coefficient values can be disaggregated in two components, one due to the net return growth and the other associated to the dividend component. The data in Table~\ref{tab:mean_alfabeta} show that the long-run return predictability is largely due to the highly predictable behaviour of the log dividend-to-price ratio. Nonetheless, especially for the case of the CRSP time series, also the net return component determines the predictive regression in a statistically validated way. The evidence that we observe confirms the empirical results discussed in Table 4 of~\cite{Campbell_Yogo:2006}. In that case it is however important to mention that the authors consider the excess gross return as dependent variable. We verified in our data that the growth in excess with respect to the long run risk free yield tends in general to better perform when we need to compute the significance of the predictive regression. In Table~\ref{tab:mean_alfabeta} we moreover show that the null hypothesis $\beta=0$ is also rejected in light of our bootstrap approach which does not rely on any ad hoc normal assumption for the idiosyncratic component $e_t$. We sample $10^4$ synthetic bootstrap copies and we measure a bootstrap $p$-value lower than 5\% for both the S\&P and NYSE/AMEX time series.\\ 
\begin{table}
  \begin{center}
    \begin{tabular}{l|rr|rr|rr}
      \rule[-7pt]{0pt}{19pt} & $\alpha$ & $\beta$ & $\alpha_\mathcal{F}$ & $\beta_\mathcal{F}$ 
      & $\alpha_\mathcal{G}$ & $\beta_\mathcal{G}$\\
      \hline
      \hline
      \rule[-7pt]{0pt}{19pt} S\&P      & 3667 & $1023 \pm 445$ & 2531 & $ 767 \pm 459 $ & $  1527 $ & $  393 \pm 19 $ \\
      \rule[-7pt]{0pt}{19pt} && (0.048) &  & (0.091) &  & $(< 10^{-3})$ \\
      \rule[-7pt]{0pt}{19pt} NYSE/AMEX & 6684 & $2228 \pm 882$ & 5681 & $1880 \pm 840$ & $ 1289 $ & $ 305 \pm 31 $ \\
      \rule[-7pt]{0pt}{19pt} && (0.024) &  & (0.043) &  & $(< 10^{-3})$ \\
    \end{tabular}
  \end{center}
  \caption{All coefficient values and associated errors are expressed on a yearly basis and in $10^{-4}$ units. Between parentheses we report the $p$-values obtained by means of $10^4$ bootstrap samples.}
  \label{tab:mean_alfabeta}
\end{table}
\begin{figure}
  \begin{center}
    \includegraphics[width=0.75\textwidth]{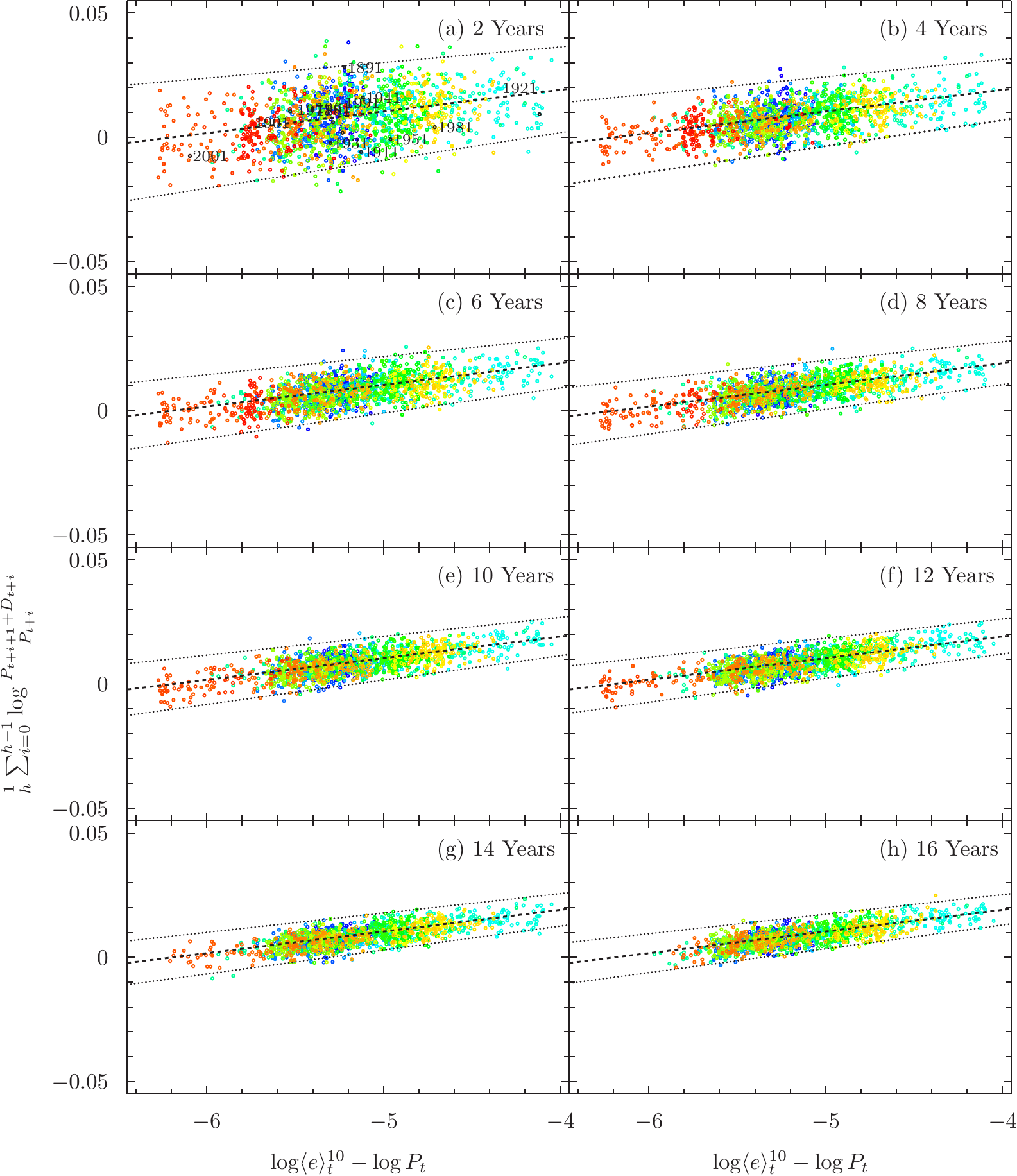}
    \caption{\label{fig:SHILLER_1871-2010_MODEL} 
    Monte Carlo scenarios generated simulating the model given by equations~(\ref{eq:linear_SDS}) and~(\ref{eq:logDP}) with inital time conditions $Y_0$, $\mu_0$, and $\log DP_0$
    equal to the empirical ones. The dashed line corresponds to the long-run behaviour predicted by the equation~(\ref{eq:mainprepo}) using the coefficient values of Table~\ref{tab:mean_alfabeta} for the S\&P index rescaled to a monthly level. The dotted lines correspond to the boundaries of the 95\% confidence level region. Points are organised in chronological order according to the color scale ranging from dark blue to red passing through light blue, green, yellow, and orange; labels in the top left panel refer to points which correspond to the first month of the specified year.}
  \end{center}
\end{figure}

\begin{figure}
  \begin{center}
    \includegraphics[width=0.75\textwidth]{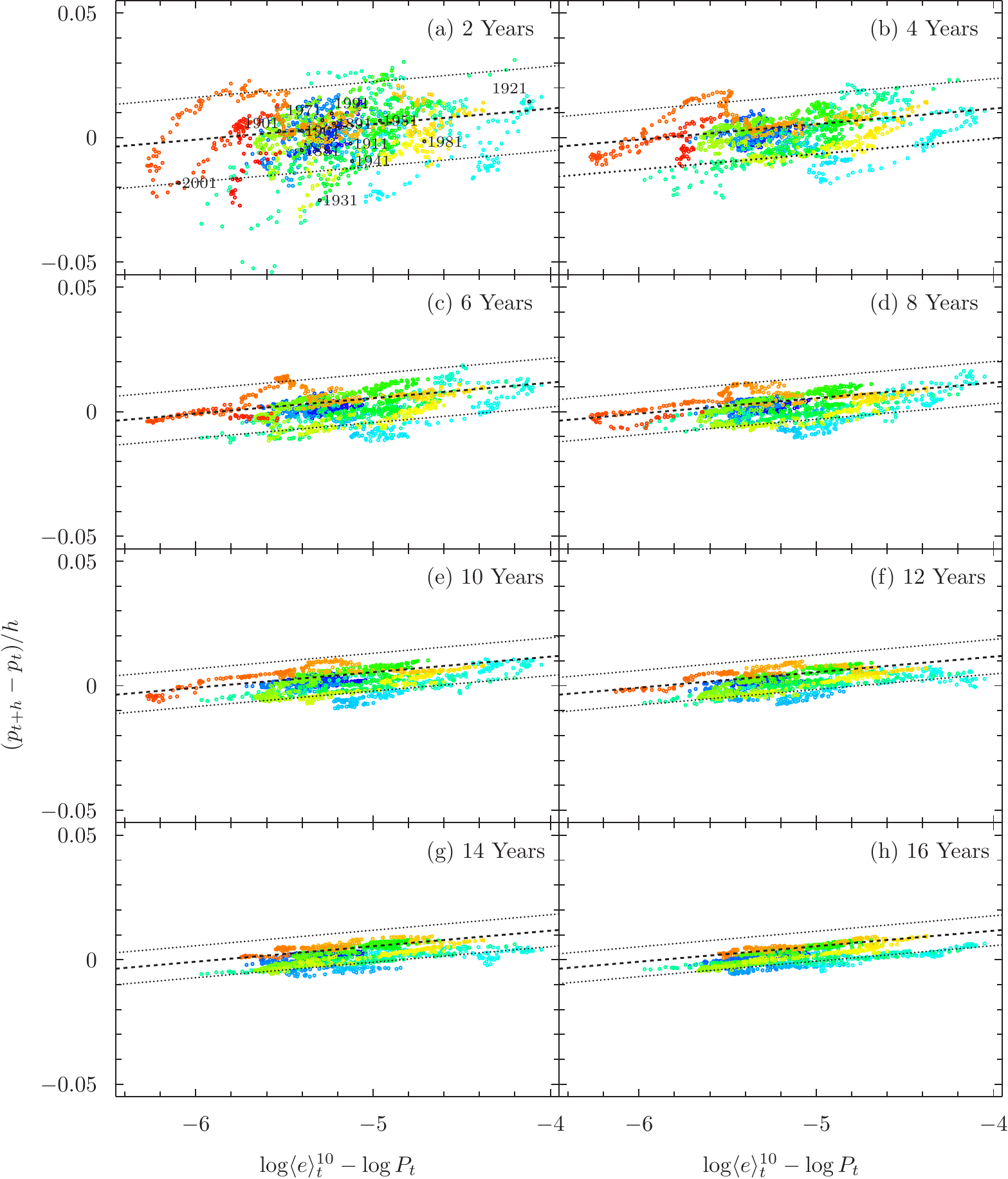}
    \caption{\label{fig:SHILLER_1871-2010_LOGP} Regression of the yield $(p_{t+h} - p_t)/h$ on the explanatory log earning - price ratio. Solid points: empirical data, dashed line: model prediction, dotted line: 95\% confidence level from model prediction. Color scale as in Figure~\ref{fig:SHILLER_1871-2010_MODEL}.}
  \end{center}
\end{figure}

\begin{figure}
  \begin{center}
    \includegraphics[width=0.75\textwidth]{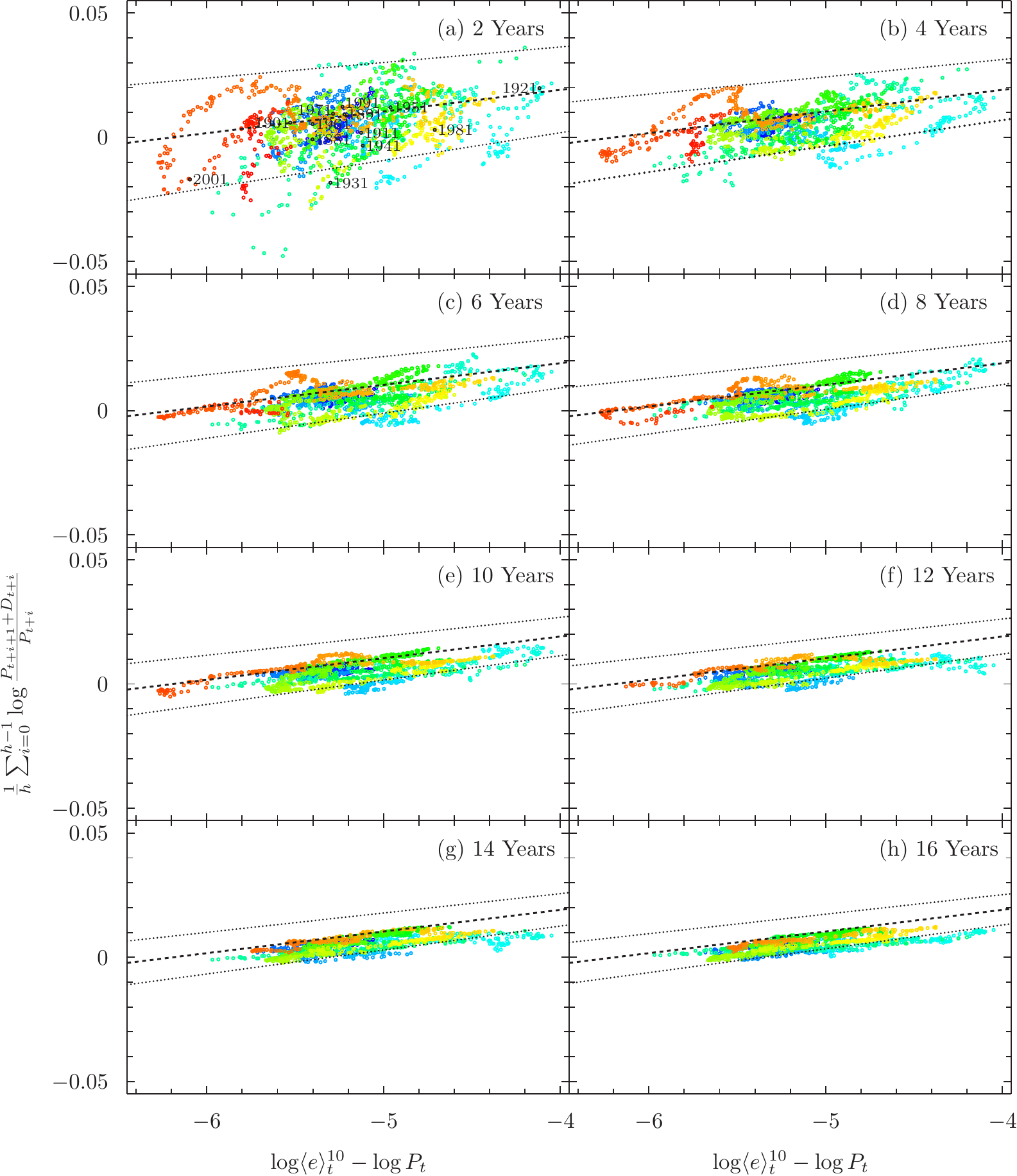}
    \caption{\label{fig:SHILLER_1871-2010_LOGDP} Regression of yield~(\ref{eq:yield}) on the explanatory log earning - price ratio. Points, lines, labels, and color scale as in Figure~\ref{fig:SHILLER_1871-2010_LOGP}.}
  \end{center}
\end{figure}

\noindent Having estimated the linear functions $\mathcal{F}$ and $\mathcal{G}$ by means of the above procedure, the next relevant parameter to estimate is the rate of exponential growth of averaged earnings $g$. This can be easily obtained iteratively, first regressing a vector of 192 monthly logarithmic averages on a linear function of $h$, then rolling the time window and repeating the estimate
from January 1881 for the S\&P index and from December 1926 for the NYSE/AMEX index until the last admissible date. In this way we obtain a sample of $g$ values from which empirical 68\% confidence intervals can be computed, for the results on a monthly basis from Shiller's and Campbell's data series please refer to Table~\ref{tab:parameters}. 
The value of $\theta$ and $\sigma_\mathrm{d}$ are estimated regressing $\Delta(d_{t-1}-p_t)$ on $d_{t-1}-p_t-\log\mathcal{G}$, employing the same technique of rolling windows. The measured mean value for $\theta$ is equal to 0.0271 and 0.0445 for the S\&P and NYSE/AMEX index respectively, and following the consideration written at the end of the previous section these values imply typical scales for mean-reversion of order 37 and 22 months, respectively. The parameters of the $\mu_t$ process are also estimated by means of linear regression of $\mu_{t+1}$ on $\mu_t$, and on $\log\langle e\rangle_t^{10}-p_t+\mathcal{H}+g\mathcal{F}t$. In the first instance this requires to introduce a suitable proxy for the unobservable variable $\mu_t$. Our approach is to estimate it by means of $p_t-p_{t-1}$, but different choices are admissible, for example in terms of an exponential weighted average of past price returns. However in this latter case the undesirable dependence of the parameters on the value of the exponential weight would be introduced. Moreover $\mathcal{H}$ is still an unknown quantity that we have to choose before performing the optimization procedure. Recalling expression~(\ref{eq:leading}), it is reasonable to assume it linear in the log EP ratio $\mathcal{H}=\alpha_\mathcal{H}+\beta_\mathcal{H}\log EP_0$ and to initialise the procedure with arbitrary values for the linear coefficients. The numerical process then produces optimal estimates of $\gamma$ and $\kappa$, for suitable $\alpha_\mathcal{H}$ and $\beta_\mathcal{H}$ minimizing the leading correction term~(\ref{eq:leading}). In particular, for $\alpha_\mathcal{H}=0.85$ and $\beta_\mathcal{H}=-0.85$ we find $\gamma=0.25$ and $\kappa=0.03$ for S\&P, and for $\alpha_\mathcal{H}=2.62$ and $\beta_\mathcal{H}=-0.52$ we find $\gamma=0.08$ and $\kappa=0.03$ for NYSE/AMEX, both satisfying the constraint $4\kappa\leq (1-\gamma)^2$. For the estimate of $\sigma_p$ we approach the problem from a different perspective. Indeed from equation~(\ref{eq:th_var}) we know that the variance of $(p_h - p_0)/h$ scales as one over $h$ with proportionality constant equal to $\sigma_p^2$. We therefore fit the empirical curves with a linear relation and we find the values of 0.182\% and 0.168\% reported in Table~\ref{tab:parameters}.
\begin{table}
  \begin{center}
    \begin{tabular}{rl|l|l}
      \hline
      \rule[-7pt]{0pt}{19pt} && S\&P & NYSE/AMEX\\
      \hline
      \hline
      \rule[-7pt]{0pt}{19pt} $g$&$(\times 10^{-4}~\textit{month}^{-1})$ & $12~(-3,31)$ & $19~(4,32)$\\
      \hline
      \rule[-7pt]{0pt}{19pt} $\theta$&$(\times 10^{-4}~\textit{month}^{-1})$ & $271~(111,430)$ & $445~(197,780)$\\
      \hline
      \rule[-7pt]{0pt}{19pt} $\gamma$ && $0.25~(0.18,0.33)$ & $0.08~(0.02,0.15)$\\
      \hline
      \rule[-7pt]{0pt}{19pt} $\kappa$&$(\times 10^{-4}~\textit{month}^{-1})$ & $323~(81,597)$ & $304~(70,547)$\\
      \hline
      \rule[-7pt]{0pt}{19pt} $\sigma^2_\mathrm{d}$&$(\times 10^{-4}~\textit{month})$ & $13~(10,20)$ & $19~(13,25)$\\
      \hline
      \rule[-7pt]{0pt}{19pt} $\sigma^2_\mu$&$(\times 10^{-4}~\textit{month})$ & $12~(9,18)$ & $17~(12,24)$\\ 
      \hline
      \rule[-7pt]{0pt}{19pt} $\sigma^2_p$&$(\times 10^{-4}~\textit{month})$ & $18.2~(18.1,18.3)$ & $16.8~(16.5,17.1)$\\
      \hline
    \end{tabular}
  \end{center}
  \caption{Parameter values for the S\&P and NYSE/AMEX time series with associated 68\% confidence level intervals.}
  \label{tab:parameters}
\end{table}
In Figure~\ref{fig:SHILLER_1871-2010_MODEL} we present the results of a Monte Carlo simulation of the model~(\ref{eq:linear_SDS})-(\ref{eq:logDP}). Each point corresponds to a single realization of $y_{t,h}$ with $h=24,\ldots,192$ months with $t$ starting from January 1881. The initial time values of $Y_0$, $\mu_0$, and $\log DP_0$ are fixed equal to the empirical ones. We plot the linear relation between yields and the log EP as predicted by the equation~(\ref{eq:mainprepo}) using the coefficient values of Table~\ref{tab:mean_alfabeta} for the S\&P index rescaled to a monthly level. We also provide the boundaries of the 95\% confidence level region which we compute by means of the formulas~(\ref{eq:th_var}) and~(\ref{eq:thvar_logdp}). 
All the analytical predictions are fully confirmed by the Monte Carlo numerical results.\\
Finally, we test the goodness of the estimated parameters and the effectiveness of our dynamical system. In Figures~(\ref{fig:SHILLER_1871-2010_LOGP}) and~(\ref{fig:SHILLER_1871-2010_LOGDP}) we plot the prediction of equation~(\ref{eq:mainprepo}) with the associated confidence band on the net returns and on the gross yields, respectively, for time horizons ranging from 24 to 192 months. As expected the model captures the shrinking of the historical data cloud with a scaling exponent dominated by the diffusive component of the price dynamics. As far as the central value is concerned, the consistency is very good for the short time horizons, whilst it worsens for longer runs. This effect is partially expected, since the linear coefficients of the predictive regressions are constants quantities which are exogenous to the dynamical model~(\ref{eq:linear_SDS}) and do not evolve with time. We are currently working on an econometric approach to the model~(\ref{eq:regression})-(\ref{eq:AR1}) in order to provide a sound theoretical basis to the extension of the low-bias procedure of~\cite{Amihud_Hurvich:2004} to long-run horizons. Preliminary results suggest that the $\beta$ coefficient rescales geometrically over time, where the basis of the geometric sequence corresponds to the AR(1) coefficient of equation~(\ref{eq:AR1}). We plan to correct for this minor effect in a future extension of our model.

\section{Conclusion and Perspectives}\label{sec:conclusions}

This paper documents predictability of stock index returns for the U.S. market on the basis of a simple valuation metric, the cyclically adjusted price-earnings ratio, and proposes a discrete time dynamical model able to capture the long-run behaviour.\\ 

\noindent Substantial evidence of the importance of fundamentals in the valuation of international stock markets has been accumulated by the proponents of fundamental indexation~\citet{Arnott_Hsu_Moore:2005}. Practitioners and academicians alike have been using several valuation measures for estimating the intrinsic value of a stock index: for example in table 2 of~\citet{Poterba_Samwick:1995} the ratio of market value of corporate stock to GDP, the year-end price-to-earnings ratio, the year-end price-to-dividend ratio and Tobin's $q$ are reported from 1947 to 1995 in an effort of alerting the reader on the possible overvaluation of the index~\note{``It is difficult to distill a simple conclusion from table 2. While price-to-earnings ratios are not unusually high at present, other measures of stock price valuation are at, or near, historical highs. ($\ldots$) Table 2 does suggest, however, that in assessing the  macroeconomic consequences of stock price movements, it may be important to distinguish between stock price fluctuations that are associated with movements in the price-to-earnings or price-to-dividends ratios and those that are not. A number of recent studies suggest that variations in the earnings-to-price ratio are correlated with prospective stock market returns, ($\ldots$). Sharp increases in either the price-to-earnings or the price-to-dividends ratio, other things equal, are associated with lower prospective returns.''}. In particular Tobin's $q$ has been proposed as another efficient method of measuring the value of the stock market, with an efficiency comparable to the CAPE~\citet{Smithers:2009}. The $q$ ratio is the ratio of price to net worth at replacement cost rather than the historic or book cost of companies. It therefore allows for the impact of inflation, much alike the CAPE which averages real earnings over a ten year span. It would be interesting to carry out an empirical analysis of the relationship between Tobin's $q$ and future stock index returns as far as to extend the present approach to countries other than the U.S.. Both perspectives are worth to be followed but require high quality long-term time series which at the moment we have not been able to find. Longer time-series would also be important so as to investigate possible ``evolutionary" phenomena like the end of the relationship between market valuation and interest rates, which may perhaps be interpreted  as an example of the adaptive market hypothesis ~\citet{Farmer_Lo:1999, Lo:2004}.

\appendix

\section{Proofs}\label{app:proof}

\begin{proof}[Proof of Lemma 1]
	The stochastic dynamical system (\ref{eq:linear_SDS}) may be put in the vector form%
	\begin{equation}
		\mathbf{V}_{t+1}=\mathbb{J}\mathbf{V}_{t}+\kappa \mathcal{H}~\mathbf{e}_2+\kappa g (1+\mathcal{F}) t~\mathbf{e}_2
		+\sigma_\mu W_{t}^\mu\mathbf{e}_2+\frac{\kappa}{1-\gamma}\sigma_p W_{t}^p\mathbf{e}_3
		\label{eq:vector_form}
	\end{equation}
	where $\mathbb{J}=\left[ 
	\begin{array}{rcc}
		1 & 1 &  1\\ 
		-\kappa  & \gamma & 0 \\ 
		0 & 0 & 1
	\end{array}
	\right] $ is the matrix of the systems, $\mathbf{e}_i$ for $i=1,2,3$ is the orthonormal canonical base,
	and $\mathbf{V}_{t}=\left[ 
	\begin{array}{c}
		Y_{t}   \\ 
		\mu_{t} \\ 
		\xi_{t}
	\end{array}
	\right] $.	
	The set of the eigenvalues of $\mathbb{J}$ includes 1, independently of $\kappa$ and $\gamma $, and two ones more  
	\begin{eqnarray*}
		\lambda_{+} &=& \frac{\gamma +1}{2}+\frac{1}{2}\sqrt{(1-\gamma)^{2}-4\kappa}\, ,\\
		\lambda_{-} &=& \frac{\gamma +1}{2}-\frac{1}{2}\sqrt{(1-\gamma)^{2}-4\kappa} \, . 
	\end{eqnarray*}
	In order to ensure $|\lambda _{+}|<1$ and $|\lambda _{-}|<1$, we need to impose the following restrictions 
	\begin{eqnarray}
		0 <\gamma <1\, ,\quad\mathrm{and}\quad  0 < \kappa \leq \frac{(1-\gamma )^{2}}{4} \, .
		\label{eq:constraints}
	\end{eqnarray}
	Iterating (\ref{eq:vector_form}) we obtain%
	\begin{eqnarray*}
		\mathbf{V}_{h}
		&=&\mathbb{J}^{h}\mathbf{V}_{0} + \kappa \left[g (1+\mathcal{F}) (h-1) + \mathcal{H}\right] \sum_{k=0}^{h-1}\mathbb{J}^{k}\mathbf{e}_{2} 
		- \kappa g (1+\mathcal{F})\sum_{k=0}^{h-1} k \mathbb{J}^{k}\mathbf{e}_{2}\\
		&&+\sigma_\mu \sum_{k=0}^{h-1}W_{k}^\mu\mathbb{J}^{h-1-k}\mathbf{e}_2+\frac{\kappa}{1-\gamma}\sigma_p\sum_{k=0}^{h-1}W_{k}^p\mathbb{J}^{h-1-k}\mathbf{e}_{3}\,.
	\end{eqnarray*}
	We can rewrite more conveniently $\mathbb{J}=\mathbb{Q}\Lambda\mathbb{Q}^{-1}$ by means of the matrices 
        $\Lambda =\left[ 
	\begin{array}{ccc}
		1 & 0 & 0 \\ 
		0 & \lambda _{+} & 0 \\ 
		0 & 0 & \lambda _{-} 
	\end{array}
	\right]$,
	$\mathbb{Q}=\left[ 
	\begin{array}{ccc}
		1-\gamma & 1 & 1 \\ 
		-\kappa & \lambda _{+}-1 & \lambda _{-}-1 \\
		\kappa & 0 & 0 
	\end{array}
	\right]$, and
	$\mathbb{Q}^{-1}=\frac{1}{|\mathbb{Q}|}\left[ 
	\begin{array}{ccc}
		0 & 0 & \lambda_--\lambda_+\\ 
		\kappa(\lambda_- - 1) &  -\kappa & (1-\gamma)(1-\lambda_{-})-\kappa\\
		\kappa(1 - \lambda_+) &   \kappa & \kappa-(1-\gamma)(1-\lambda_{+})
	\end{array}
	\right]$
	with $|\mathbb{Q}|=\kappa(\lambda_--\lambda_+)$.
	Taking expectation of $\mathbf{e}_1^\mathrm{t} \mathbf{V}_h$ we obtain
	\begin{eqnarray}
		\mathbb{E}_{0}\left[ Y_{h}\right] &=& \mathbf{e}^\mathrm{t}_1\mathbb{Q}\Lambda^{h}\mathbb{Q}^{-1}\mathbf{V}_0\notag\\
		&&+\kappa \left[g (1+\mathcal{F}) (h-1) + \mathcal{H}\right]\sum_{k=0}^{h-1}\mathbf{e}^\mathrm{t}_1\mathbb{Q}\Lambda^{k}\mathbb{Q}^{-1}\mathbf{e}_{2}\notag\\
		&&-\kappa g (1+\mathcal{F})\sum_{k=0}^{h-1} k \mathbf{e}^\mathrm{t}_1\mathbb{Q}\Lambda^{k}\mathbb{Q}^{-1}\mathbf{e}_{2}\notag\\
		&=&-\frac{\lambda_+^h}{\lambda_- - \lambda_+}\left[(1-\lambda_-)Y_0+\mu_0\right]
		+\frac{\lambda_-^h}{\lambda_- - \lambda_+}\left[(1-\lambda_+)Y_0+\mu_0\right]\notag\notag \\
		&& + g (1+\mathcal{F}) (h-1) + \mathcal{H}\notag\\
		&& - \left[g (1+\mathcal{F})-\mathcal{H}\right]\frac{\kappa}{\lambda_- - \lambda_+}\left(\frac{\lambda_+^h}{1-\lambda_+}-\frac{\lambda_-^h}{1-\lambda_-}\right)\notag\notag\\
		&& - g (1+\mathcal{F})\frac{\kappa}{\lambda_- - \lambda_+}\left[\lambda_-\frac{1-\lambda_-^h}{(1-\lambda_-)^2}-\lambda_+\frac{1-\lambda_+^h}{(1-\lambda_+)^2}\right]\, .
		\label{eq:expected_Y}
	\end{eqnarray}
	From previous expression relation~(\ref{eq:th_mean}) immediately follows.\\
	As far as the variance of $Y_{h}$ is concerned, we have 
	\begin{equation}
		Var_{0}\left[ Y_{h}\right] 
		= \underset{k=0}{\overset{h-1}{\sum }}\left[\sigma_\mu^{2}\left( \mathbf{e}_{1}^{\mathrm{t}}\mathbb{Q}\Lambda ^{k}\mathbb{Q}^{-1}\mathbf{e}_2\right)^{2}
		+\frac{\kappa^2}{(1-\gamma)^2}\sigma_p^{2}\left( \mathbf{e}_{1}^{\mathrm{t}}\mathbb{Q}\Lambda ^{k}\mathbb{Q}^{-1}\mathbf{e}_{3}\right)^{2}\right]\, ,
	\end{equation}%
	from which we find 
	\begin{eqnarray*}
		Var_{0}[Y_{h}]&=&
		\frac{\sigma_\mu^{2}}{(\lambda_- - \lambda_+)^2}\sum_{k=0}^{h-1}\left[\lambda _{-}^{k} - \lambda_{+}^{k}\right]^{2}\\
		&&+\frac{\sigma_p^{2}}{(\lambda_- - \lambda_+)^2}\sum_{k=0}^{h-1}\left[\lambda_- - \lambda_+
		+\lambda_{+}^{k}(1-\lambda_{-}-\frac{\kappa}{1-\gamma})\right.\\
		&&\left.+\lambda_{-}^{k}(\frac{\kappa}{1-\gamma}-1+\lambda_{+})\right]^{2}\, ,
	\end{eqnarray*}
	and the thesis~(\ref{eq:th_var}) follows.
\end{proof}
\begin{proof}[Proof of Lemma 3]
  On a monthly basis the quantity $D_{t}/P_{t+1}$ is of order $10^{-3}$, and we are allowed to replace $\log{\left(1+\frac{D_{t}}{P_{t+1}}\right)}$
  with $D_{t}/P_{t+1}$.  
  In the same spirit of~\cite{Campbell_Shiller:1988a,Campbell_Shiller:1988b}, in order to prove the thesis we need to Taylor 
  expand the quantities of interest around $\log \mathcal{G}$. In particular, we have
  \begin{equation*}
    \frac{D_{t}}{P_{t+1}} = \exp{\left(d_{t} - p_{t+1}\right)}\simeq \exp{\left(\log \mathcal{G}\right)}
    \left(1+d_{t} - p_{t+1}-\log\mathcal{G}\right)\, .
  \end{equation*}
  Iterating equation~(\ref{eq:logDP}) and taking expectation, we obtain for $t\geq 0$
  \begin{equation}
    \mathbb{E}_0[d_{t-1}-p_t] = (1-\theta)^t (d_{-1}-p_0) + \log\mathcal{G}\left[1-(1-\theta)^t\right]\, ,
    \label{eq:expected_dX}
  \end{equation}
  where $d_{-1}$ is the log dividend prevailing at time zero being cumulated between time -1 and zero. Under the constraint $0<\theta<2$, we can conclude
  \begin{eqnarray*}
    \frac{1}{h}\mathbb{E}_0\left[\sum_{i=0}^{h-1} \log{\left(1+\frac{D_{i}}{P_{i+1}}\right)}\right]&\simeq& \mathcal{G}(1-\log\mathcal{G})+\mathcal{G}\log\mathcal{G}\\
    &&+\mathcal{G}\frac{1-\theta}{\theta}(\log DP_0-\log\mathcal{G})\frac{1-(1-\theta)^h}{h}\\
    &=&\mathcal{G}+O\left(\frac{1}{h}\right)\,,
  \end{eqnarray*}
  and
  \begin{eqnarray}
    Var_0\left[\frac{1}{h}\sum_{i=0}^{h-1} \log{\left(1+\frac{D_{i}}{P_{i+1}}\right)}\right]&\simeq& \frac{1}{h}\frac{\mathcal{G}^2 \sigma_\mathrm{d}^2}{\theta(2-\theta)}+o\left(\frac{1}{h}\right)\,.
    \label{eq:thvar_logdp}
  \end{eqnarray}
\end{proof}
\noindent\textit{Remark}. The expansion performed in the above proof is similar to that proposed in~\cite{Campbell_Shiller:1988a,Campbell_Shiller:1988b}, where $\log (P_t + D_t)$ is approximated by $\vartheta p_t + (1-\vartheta) \log D_t + \varkappa$ with $\vartheta=1/(1+\mathcal{G})$ and 
$\varkappa=\log (1+\mathcal{G})-\vartheta \mathcal{G} \log\mathcal{G}$. 

\end{document}